\documentstyle[11pt,newpasp,twoside,psfig]{article}
\newbox\grsign \setbox\grsign=\hbox{$>$} \newdimen\grdimen \grdimen=\ht\grsign
\newbox\simlessbox \newbox\simgreatbox
\setbox\simgreatbox=\hbox{\raise.5ex\hbox{$>$}\llap
     {\lower.5ex\hbox{$\sim$}}}\ht1=\grdimen\dp1=0pt
\setbox\simlessbox=\hbox{\raise.5ex\hbox{$<$}\llap
     {\lower.5ex\hbox{$\sim$}}}\ht2=\grdimen\dp2=0pt

\def\simless{\mathrel{\copy\simlessbox}}
\def\etal{{\it et al.\ }}
\def\eg{{\it e.g.,\ }}

\def\edcomment#1{\iffalse\marginpar{\raggedright\sl#1\/}\else\relax\fi}
\reversemarginpar

\begin{document}
\title{Multi-wavelength Surveys for Distant Clusters}
\author{Marc Postman}
\affil{Space Telescope Science Institute, 3700 San Martin Drive,
Baltimore, MD 21208, U.S.A.}

\begin{abstract}
Given the importance of clusters to the fields of cosmology and
galaxy evolution, it is critical to understand how the cluster
detection process affects (biases) ones scientific conclusions
derived from a given cluster sample. I review the astrophysics, algorithms,
and observational constraints that must be considered when
attempting to assess cluster selection biases and their impact on the
scientific constraints derived from multi-wavelength cluster surveys. 
In particular, one can more accurately quantify and understand the
selection biases and assess cluster evolution 
when using joint optical/NIR/x-ray surveys than can be achieved when
employing any single cluster survey.
\end{abstract}

\section{Introduction}

It is well appreciated by this audience that clusters of galaxies
are important as tracers of mass on intermediate and large scales
and as laboratories for exploring galaxy evolution.
Recently, the application of coordinated
multi-wavelength observations has significantly enhanced our understanding
of the physics of cluster formation and evolution. In particular,
the use of multi-wavelength data
\begin{itemize}
\item enables studies of a broad range of physical processes which control
cluster and cluster galaxy evolution,
\item expands the accessible redshift range,
\item enhances detection of distant large-scale structure (e.g., intercluster
filaments) via joint passband detection methods, and
\item minimizes false positive detections and improves ones understanding
of sample selection biases.
\end{itemize}
Now that the era of objective cluster detection across wide wavelength
ranges is upon us, it is critical to ask {\bf ``How does the cluster detection
process affect (bias) the scientific conclusions?"} This is an important
question because {\it different types of clusters} can be found by
different algorithms and/or at different wavelengths. The different
detection outcomes are a consequence of a cluster detection S/N ratio
that is generally a function of redshift, survey passband, and the mathematical
details and statistical properties of the detection algorithm applied.

The wavelength-dependent differences in the properties of clusters are
directly related to the physical phenomena that dominate the clusters 
``visibility." In the x-ray passbands (0.1 -- 10 keV), it is the hot
intracluster medium (ICM) that is being detected. The ICM in clusters
typically shines with a luminosity in the range $43 < {\rm log(L_x)} < 45.5$.
A luminous, centrally condensed ICM usually indicates a cluster in
a state of dynamical equilibrium with a fairly deep gravitational potential.
In the optical and near-IR passbands (0.6 -- 2.2$\mu$), it is one or more of the
following that is being detected: a large overdensity of galaxies, a population
of early-type galaxies with a narrow optical/NIR color range, and/or the
distortion of the images of background galaxies due to gravitational lensing
by the cluster. A large galaxy overdensity does not necessarily imply
a well-relaxed or massive cluster. Hence, optical/NIR based surveys may include systems
of significantly lower mass and/or age than those found in x-ray based surveys.
The spectral energy distribution of most clusters peak in the optical/NIR
with luminosities in the range $44 < {\rm log(L_{opt})} < 46.5$.
The millimeter wavebands (200 -- 300 GHz) are ideal for detection of
the Sunyaev-Zeldovich effect. 
The detection of the SZ effect is independent of redshift
in as much as the electron density in the ICM and the effective electron temperature is
independent of redshift (i.e., if $n_e$ and $T_e$ are functions of time, then
so will be the amplitude of the SZ decrement). SZ based surveys have great potential for 
detecting systems in a quite homogeneous manner out to $z > 1$. However, the
presence of a significant ICM is obviously required. At the longest wavelengths
(1.4 -- 30 GHz), detection of radio loud cluster galaxies at very high redshifts
and detection of galaxies with bent radio lobes (suggesting passage through
an ICM) can be achieved. Because radio-loud cluster galaxies may not always
be a by-product of cluster formation, radio-based cluster surveys have
significant selection biases. They are, nonetheless, capable of detecting systems
at the highest redshifts ($z > 2$).

\section{Astronomical \& Observational Considerations}

In addition to a variety of astrophysical parameters that dictate
the area, depth, and geometry of a distant cluster survey, there are also 
several important observational considerations that must
be accounted for. An obvious one is the much greater availability of
optical/NIR facilities relative to x-ray telescopes. This makes completing
large area surveys in the optical/NIR much easier. Competing with this
is the fact that the detection of the ICM is a relatively unambiguous indicator of
the physical reality of the cluster candidate. The pros and cons of performing
cluster surveys in x-ray, optical, and SZ bandpasses are summarized in the
table below.  

The lack of a significant background in the x-rays is a key advantage of
looking for clusters in this wavelength range. Figure~1 shows a beautiful
example of this in the form of Chandra/ACIS imaging of the 
Lynx supercluster complex at $z \sim 1.3$
(Stanford \etal 2001). However, such x-ray images require a large
investment of telescope time (190 ksec in the example here). The same
clusters can be easily detected in NIR images with ${1 \over 10}$th the
exposure time. In addition,  the presence of
a significant background in the optical/NIR when searching for clusters
at intermediate and high redshifts has lead to the development of several
smart algorithms which selectively suppress the background and, in doing
so, yield relatively accurate ($\vert \delta z\vert/z \simless 0.1$) redshift estimates
for the cluster candidates. Having these redshift estimates greatly accelerates
the science that can be extracted from a survey conducted primarily\footnote{I
say primarily because the use of color information can provide a quasi-third
dimension that eliminates projection effects on scales larger than $\sim10,000$
km s$^{-1}$.} in two dimensions. Note that the $z=1.27$ Lynx cluster in Figure~1,
CL0848+4453, was initially discovered in a NIR image (Stanford \etal 1997).
\begin{figure*}
\centering\mbox{\psfig{figure=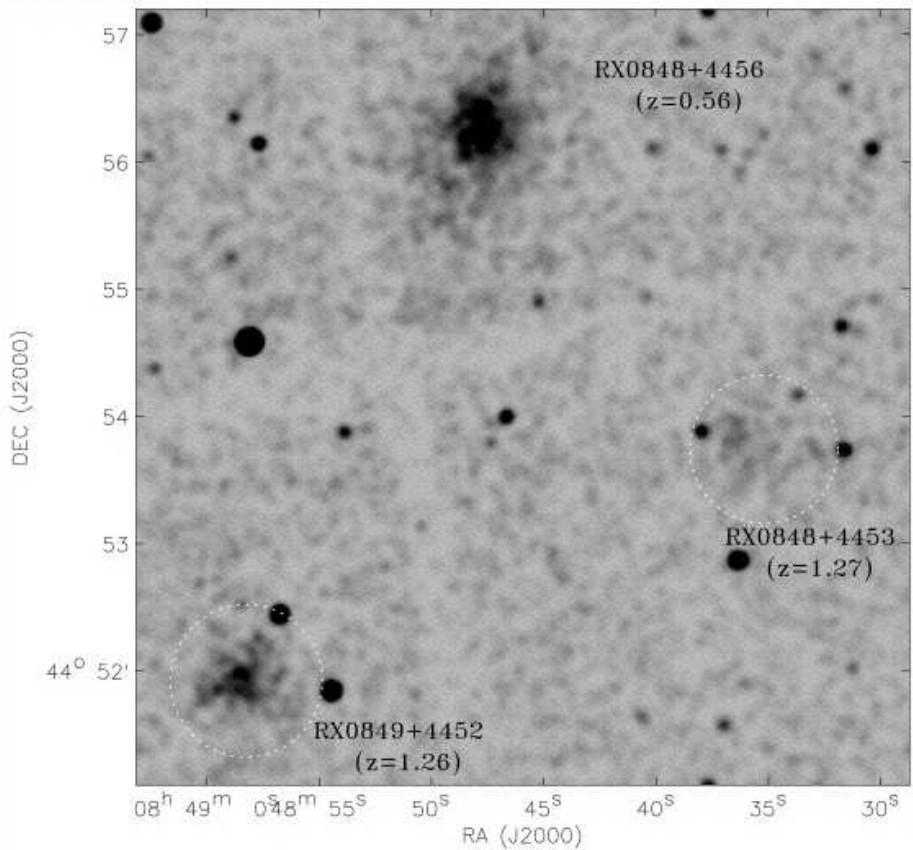,height=4.5in}}
\caption[]{A 190 ksec Chandra/ACIS image of the Lynx supercluster at 
$z=1.27$. Figure taken from Stanford \etal (2001).}
\end{figure*}

Figure~2 summarizes the cluster survey area--depth parameter space
covered to date (with a few near-future surveys included). This figure is modeled
after that in Ebeling et al. (2001) but includes optical and NIR surveys
as well. Survey flux limits
have been converted to common cgs units for comparison across a broad wavelength
range. Although the primary science motive for each listed survey is not 
always the study of cluster evolution, each one is well suited to the search
for distant clusters.  Within a given passband, the survey flux limit is
an indicator of the effective redshift depth of the survey. However, in order to
directly compare effective depths across a large spectral range one must
also understand how the cluster and galaxy luminosity functions vary 
with wavelength. Fortunately,
this is relatively well known. The
characteristic luminosity of clusters in the x-ray (0.5 -- 2 keV), 
and the characteristic luminosity of cluster galaxies in the optical
(0.8$\mu$), and NIR (2.2$\mu$) passbands are plotted for $z = 0.15$ and $z = 1$.
Also shown are curves which indicate the areas and depths at which one
would expect to find 10 and 100 clusters, respectively, with $L_x > 5 \times 10^{44}$
erg s$^{-1}$ at any z (Ebeling \etal 2001).  
\begin{figure*}
\centering\mbox{\psfig{figure=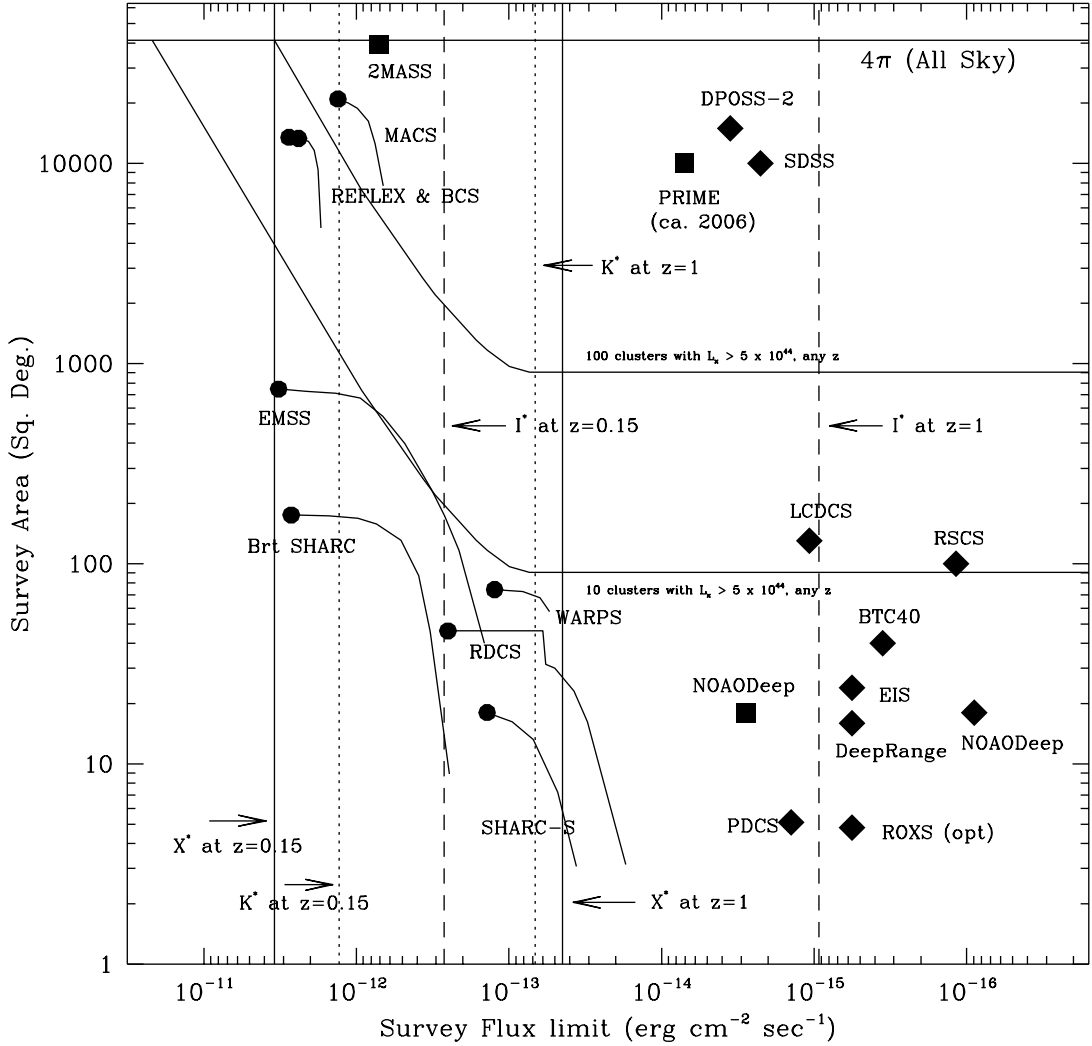,height=6in}}
\caption[]{Survey flux limit versus survey area for a range of sky survey
projects either completed or slated for completion in the near future.
X-ray surveys are shown as large circles with
trailing curves, to reflect the decreasing sensitivity of x-ray telescopes
as targets move off-axis. NIR surveys are shown as squares, optical surveys
are shown as diamonds.
Vertical lines are the characteristic luminosities of clusters in the x-ray
(0.5 -- 2 keV) and of cluster galaxies in the optical (0.8$\mu$) 
and NIR (2.2$\mu$) passbands at $z = 0.15$ and $z = 1$.
Data points in this figure are adopted from
information provided in Postman \etal 1996; Postman \etal 1998;
Jannuzi et al. 2000; de Propris et al. 1999; Voit \& Donahue 1999;
and Ebeling et al. 2001}
\end{figure*}

\begin{table}
\begin{tabular}{l|l|l}
\tableline
\tableline
X-ray Surveys & Optical/NIR Surveys & SZ (mm) Surveys \\
\tableline
Detection of ICM is an  & Large areas can be surveyed & Detection of decrement \\
unambiguous indicator of & relatively quickly          & requires ICM, an \\
a real cluster          &                             & unambiguous indicator of \\
                        & Many telescopes available   & a real cluster \\
High contrast $(L_x \propto n_e^2)$ &                 & \\
                        & High QE detectors           & SZ signal is $z$ independent\\
Background is negligible&                             & (modulo ICM evolution)\\
                        & Many algorithms             &                         \\
Low (10\%) spurious rate& produce cluster redshift    & Limited telescope access\\
                        & estimate                    & \\
Limited telescope access&                             & Large area surveys are time \\
                        & 30\% spurious rate (in 2D)  & intensive but upgrades to \\
Large area surveys are  &                             & BIMA/OVRO could cut \\
time intensive          & Background is significant,  & time by factor of 10 \\
                        & need ``smart" algorithms    & \\
Requires optical/NIR    &                             & \\
follow-up for redshifts & Optical richness - cluster  & \\
                        & mass relation is noisy      & \\
\tableline
\end{tabular}
\end{table}

Recent deep optical/NIR surveys are now reaching areas in excess of
100 square degrees enabling the discovery of the richest clusters at $z \sim 1$
and the study of the evolution of 
the moments of the cluster distribution. In this regard, optical/NIR surveys
still provide the largest numbers of distant ($z > 0.6$) cluster candidates.
However, new untargetted SZ surveys and very wide area x-ray surveys
(\eg MACS) provide substantial potential for the creation of distant
cluster catalogs with very low contamination rates.
The NOAO Deep Wide survey (Jannuzi et al. 2000)
and the proposed {\it Primordial Explorer} (see http://prime.pha.jhu.edu)
mission will provide superb datasets, enabling detection and
study of clusters out to $z \sim 2$. 

An optimal cluster survey strategy would be to take advantage of
the benefits of each particular passband by observing a given
region of sky in optical/NIR and x-rays. At the least, this will yield
important constraints on differences in the selection functions at these
wavelengths. At best, this will produce a comprehensive set of data with
which to study cluster evolution. While this is currently difficult to
do to sufficient depth to study distant clusters over a large area of sky,
it is feasible over regions of a few square degrees. Indeed, several teams have recently
done precisely such studies either using existing x-ray data and re-imaging
in the optical (Donahue \etal 2001) or using existing optical data and
searching in x-ray archives for the corresponding high energy imaging
(Holden \etal 1999). I review the conclusions reached from these
two studies later in this talk to demonstrate the important constraints
that multi-wavelength distant cluster surveys can yield.
First, however, I will briefly review the cluster detection algorithms in use
today, emphasizing some of the wavelength dependent pros and cons of each, 
as well as highlighting the importance of quantifying the selection function
prior to attempting scientific analysis of any objectively derived cluster
sample.
 
\section{Cluster Detection Algorithms}

The advancement of cluster detection algorithms has benefited greatly from
the adaptation of well-known signal and image processing techniques to
astronomical use. These
include matched filters (\eg Postman \etal 1996, Kawasaki \etal 1998,
Kepner \etal 1999, Lobo \etal 2000, Gladders \& Yee 2000) and adaptive
wavelet transforms (\eg Rosati \etal 1995). Other methods come to astronomy
from the realm of multi-parametric statistical and clustering analysis such as the
Voronoi Tessellation Technique (\eg Scharf \etal 1997, Ramella \etal 2001).
One characteristic
of distant clusters, almost regardless of observational bandpass, is that
they are relatively low contrast features\footnote{The contrast of a distant cluster
in a survey is often low because the survey goal is to cover a modest to large
area at the expense of depth. The contrast of the same cluster in a given 
bandpass can, of course, be increased by increasing the exposure time as in 
the Chandra/ACIS example in Figure~1.}.
A key property of many of the above
methods is, thus, to take the basic input data (\eg, an x-ray image or a galaxy
catalog) and process it in a manner that significantly enhances the contrast.
For example, the matched filter algorithm developed by Postman \etal (1996),
and modified further by Kepner \etal (1999), enhances the contrast
of a $z \sim 0.8$ cluster observed in the $I-$band sufficiently to turn a 
$2.4\sigma$ fluctuation in raw galaxy counts into a $4.5\sigma$ fluctuation
in the matched filter signal. Similarly, the color cut used by 
Gladders \& Yee (2000) in their red-sequence detection method can transform
a $2\sigma$ fluctuation in galaxy counts (due to a $z \sim 1$ cluster)
into a $5\sigma$ fluctuation. Most of the methods above can be applied
at a wide range of redshifts and over a broad spectral range. Exceptions include
methods which rely on cluster characteristics that span a limited spectral
range (\eg the red galaxy sequence) or are best applied in the optical/NIR
(\eg detection of gravitational lensing). Because each method comes with 
advantages and disadvantages, using multiple methods is desirable. Multiple
algorithm application can assure high completeness across a broad range of
cluster properties. 
\begin{figure*}
\centering\mbox{\psfig{figure=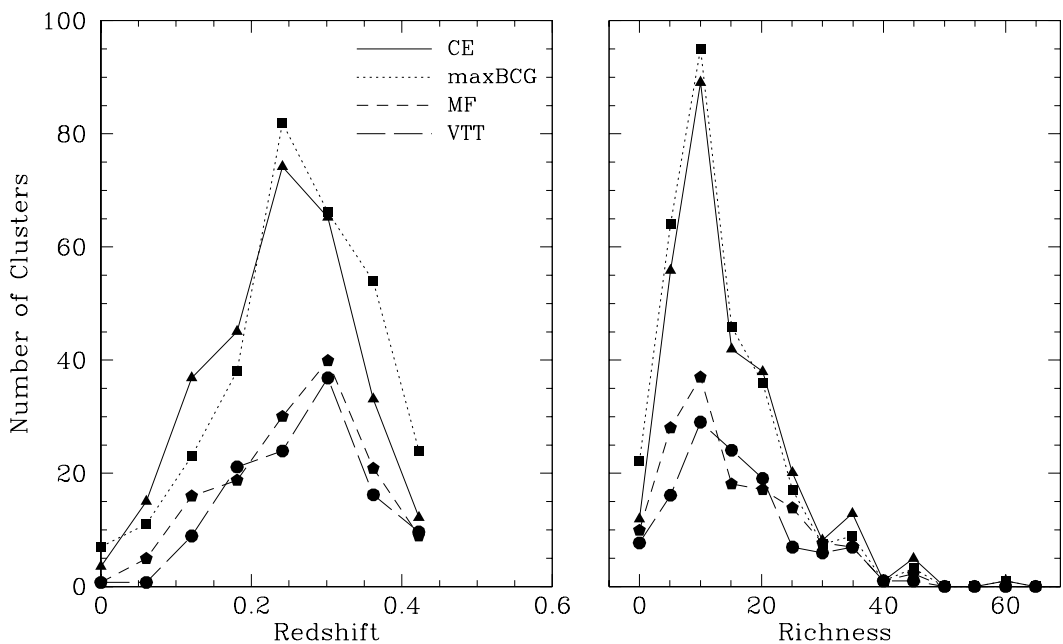,height=4.5in,angle=270}}
\caption[]{Number of clusters detected as functions of redshift
and richness for 4 different search algorithms. The input data to
each algorithm was identical and is taken from the SDSS early data
release. This figure is adopted from the work of Goto \etal 2001.
The 4 detection methods are: CE (cut \& enhance algorithm which
weights color information in a special way to increase the contrast
of a cluster), maxBCG (searches for the red sequence and the BCG),
MF (Matched filter algorithm), VTT (Voronoi Tessellation Technique).
The histograms have NOT been corrected for the different selection
functions intrinsic to each algorithm.}
\end{figure*}
 
Indeed, the use of multiple algorithms highlights the importance of 
determining an accurate selection function. Figure~3, adopted from the
work of Goto \etal 2001, shows an example
of the differences in the richness and redshift distributions of cluster
candidates identified using 4 different detection methods applied to
identical data. The input data was the early release data from the 
Sloan Digital Sky Survey. The differences in the resulting cluster catalogs
can be understood as a consequence of the different selection
function each algorithm yields. Accurate interpretation of cluster data is thus
critically dependent on accurate estimation of the relevant selection
function.
 
An example of how cluster selection might affect scientific conclusions
is a comparison of the recent work of Dressler \etal 1999 and 
Poggianti \etal 1999 (MORPHS) versus that of Balogh \etal 1999 
and Ellingson \etal 2001 (CNOC). 
The CNOC study used 15 x-ray selected ($L_x > 2 \times 10^{44}$ erg s$^{-1}$)
clusters in the redshift range $0.18 < z < 0.55$. These clusters are 
mostly rich, relaxed systems. The CNOC clusters also exhibit a prominent
E/S0 red sequence in their color-magnitude diagrams. The 10 MORPHS clusters
were primarily optically selected and lie in a similar redshift range,
$0.37 < z < 0.56$. They span a wide range of mass and $L_x$, with 40\% being
classified as ``irregular" in their overall morphology (as traced by the
galaxy distribution). The CNOC clusters are reported to contain a low
$k+$A galaxy fraction of 4.5\% and after correction for scatter this value
decreases to 1.5\%. The field $k+$A fraction, for comparison, is 1 -- 2\%.
In contrast, the MORPHS survey finds a significantly larger $k+$A fraction
of $\sim20$\% and a significant excess of post-starburst galaxies. The MORPHS
study also concludes that star formation in clusters is truncated relatively
quickly ($\sim1$ Gyr). The CNOC study finds no excess starburst or post-starburst
activity relative to the field and concludes that cluster galaxy star formation
is truncated gradually over a 2 -- 3 Gyr period. How can these two studies
reach such different conclusions given the similarity in sample size and
redshift range? The answer may, in part, lie in the different selection 
biases associated with optical vs. x-ray selection. X-ray luminous clusters
at $z \simless 0.5$ tend to be dynamically older systems -- as indicated by
the existence of a well-established and centrally-compact intracluster medium. 
Their galaxy populations, at least within the central 500$h^{-1}$ kpc, are
dominated by early type galaxies with a narrow range in optical color. 
While such systems would easily be detected in optical surveys, systems 
that are dynamically younger may often not satisfy the minimum x-ray 
luminosity constraint and would thus be systematically excluded from an x-ray 
selected survey. Infall of late type galaxies into clusters may be more prominent 
in younger clusters and, hence, so to the presence of starburst 
and post-starburst objects. While one cannot attribute all differences between
the CNOC and MORPHS survey results to the cluster selection process (\eg some
of the differences are likely due to differences in the data analysis or data quality),
it is clearly an important factor.

\section{Case Studies}

To better understand the consequences of using x-ray and optical selection criteria,
as well as to improve constraints on cluster evolution in general, it is essential
to study a common region of sky in multiple wavelengths. Two recent works, the
CFHT Optical PDCS Survey (COPS; Holden \etal 1997, 1999, 2000; Adami \etal 2000)
and the ROSAT Optical X-ray Survey (ROXS; Donahue \etal 2001), have done just this
and have reached some interesting conclusions which are worth reviewing here.

In the COPS study, the ROSAT archive was searched for PSPC data with
exposure times of 3 ksec or more and lying within 40
arcminutes of the centers of 31 optically selected clusters from the Palomar
Distant Cluster Survey. The x-ray data were then processed using the SHARC
source detection software (Romer \etal 2000). A cumulative x-ray luminosity
function (XLF) was derived from the resulting extended source catalog and compared
with previous XLFs from Burns \etal (1996), Ebeling \etal (1997), Burke \etal (1997),
and Rosati \etal (1998). Of the 31 PDCS clusters studied, 7 (23\%) were detected
in the x-ray data. The cumulative XLF of both the detections and the upper limits
are consistent with those from previous x-ray selected samples. Holden \etal conclude
that {\it optical selection does not appear to miss a significant number of
x-ray emitting clusters within a given area} and that optical selection finds
both intrinsically luminous and intrinsically faint ($L_x \le 10^{43}$ erg s$^{-1}$)
x-ray clusters. Of course, the errors on the derived XLF in the COPS study are
substantial given the small sample. Further, while this effort was the first to
systematically explore objectively derived optical and x-ray cluster samples it
was not a double blind experiment -- the existence of an optically selected cluster
was the trigger for the ROSAT archive search.

The ROXS study takes the spirit of the COPS study and raises it 
one level higher by performing
completely independent optical and x-ray cluster detection over a common 4.8 deg$^2$
area. The ROXS x-ray data are 23 deep ($T_{exp} >$ 8 ksec) ROSAT PSPC pointings.
Optical images in the $V$ and $I$ passbands were obtained for each ROSAT pointing.
An adaptive wavelet algorithm (Rosati \etal 1995) was applied to the ROSAT images
to detect extended sources. A matched filter algorithm was applied to the optical
data to detect clusters and derive an optical richness estimate, $\Lambda_{CL}$,
which corresponds to an effective optical luminosity. 
The availability of complete x-ray and optical data over
the same sky area enabled the ROXS team to generate entirely independent cluster
catalogs. The two catalogs were compared in a variety of ways and the following key
results derived:
\begin{itemize}
\item 57 x-ray detections and 155 optical detections were found. 
72\% of the x-ray candidates
are co-identified in the optical survey. Of the remaining 28\%, one third are bona fide
optically faint candidates. The rest are systems with poor x-ray flux measurements.
\item Most $\Lambda_{CL}>30$ optically selected clusters are real {\it if} the
$L_x \propto \Lambda_{CL}^{\beta}$ relationship is steep. The best fit is $\beta=3.8\pm0.8$.
\item {\it There is no need to hypothesize a sizeable population of optically rich,
x-ray faint clusters}. The observed cluster distribution is consistent with a steep
$L_x - \Lambda_{CL}$ relation in which there is also significant scatter between x-ray
and optical luminosity and a false positive rate (in the optical) of $\sim25$\%.
\end{itemize}
Indeed, if the slope of the $L_x - \Lambda_{CL}^{\beta}$ relation is steeper
than 2 and $\Lambda_{CL} \propto L_{Opt}$, then the $M/L$ ratio of clusters will
continue to increase with mass (at least into the moderate mass range of ROXS
clusters). Specifically, if $L_x \propto M^2$ then $M/L_{Opt} \propto M^{1-2/\beta}$.
Figure~4 shows this prediction overplotted on actual observational data obtained
for low redshift clusters (Hradecky \etal 2000).
\begin{figure*}
\centering\mbox{\psfig{figure=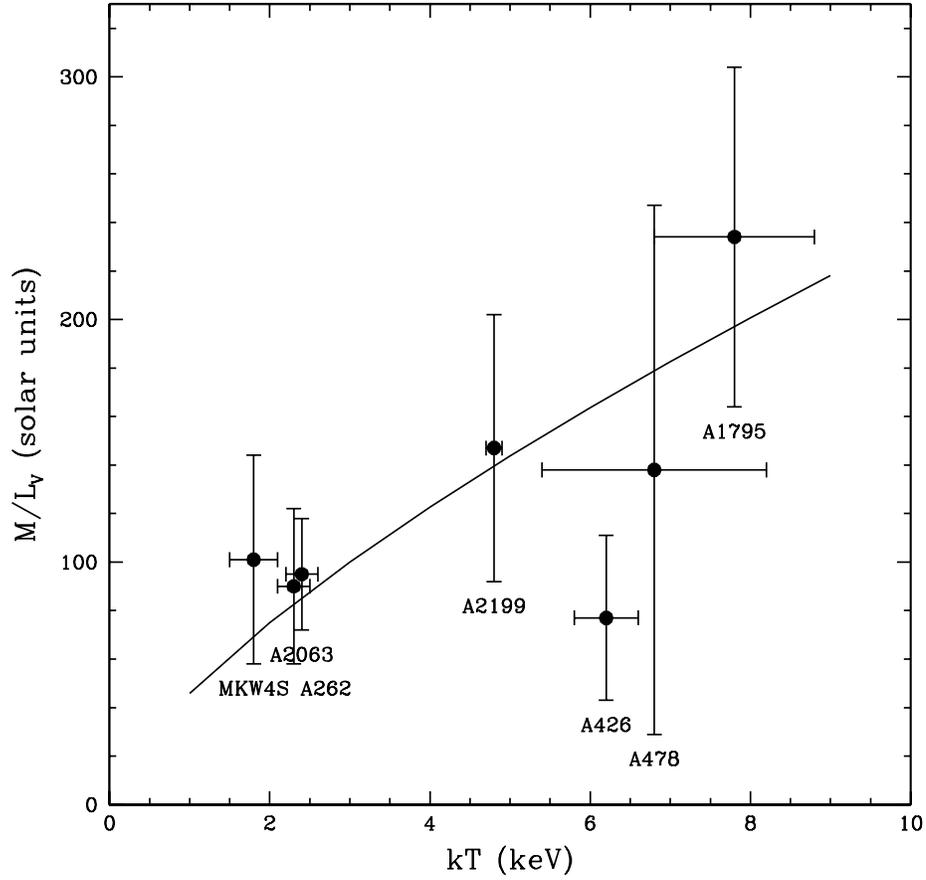,height=5in}}
\caption[]{The optical M/L ratio of clusters as a function of the
kinetic gas temperature of the ICM. Data points are from Hradecky \etal 2000.
The solid curve is the prediction from the relationship
$M/L_{Opt} \propto M^{1-2/\beta}$. A $L_x \propto T_x^{2.8}$ relation
is adopted to convert luminosity to temperature.}
\end{figure*}

\section{Goals for Multi-wavelength Distant Cluster Surveys}

Multi-wavelength surveys for distant clusters have conclusively demonstrated their
value in quantifying wavelength-dependent selection biases, establishing 
fundamental relationships between global cluster properties, and for
predicting and characterizing key evolutionary trends. The near term goals for
the current and next generation of cluster surveys will likely focus on addressing
the following key scientific questions:
\begin{itemize}
\item How do global cluster properties (mass, gas fraction, mass profile) depend
on the size of the early type galaxy population, star formation rate, and internal
kinematics. For example, {\it is the red sequence unique to virialized clusters?}
\item How does clustering on large-scales evolve? Cluster surveys are now reaching
volumes and depths that enable detection of several hundred clusters with
$z > 0.6$. The utility of clusters as tracers of large-scale structure has been
proven at low redshift and that same advantage can now be employed at $z \sim 1$.
\item How do clusters themselves evolve? The assembly of a statistically complete
sample of clusters in the range $1 < z < 2$ over an area of $>100$ deg$^2$ is
within reach of current technology (NIR mosaic cameras, more efficient millimeter
band receivers for SZE detection). Coupled with spectroscopic, HST, and x-ray observations,
such a large survey will yield precise constraints on cluster evolution.
\end{itemize}
To achieve these goals, equally dedicated efforts must be undertaken in 
the development and/or enhancement of cluster detection algorithms
(\eg, see Nichol \etal 2001) and improved methods for comparing and analyzing
independently derived cluster catalogs over a broad spectral range. 
This is especially important given the current or near future
availability of homogeneous multi-wavelength
surveys that have significant sky overlap (\eg 2MASS, SDSS, ROSAT/MACS, MAP, PRIME).
The LSST (Tyson, Wittman, \& Angel 2000) 
holds the potential for wide-area ``mass"-selected cluster catalogs,
substantially simplifying the scientific interpretation of the resulting cluster samples. 
In all these efforts, however, the question of understanding how the cluster detection
process affects the scientific conclusions must always be in the fore.

\end{document}